\newif\ifFull
\Fulltrue
\newif\ifIEEE
\IEEEfalse

\ifIEEE
\documentclass[10pt,journal,letterpaper,compsoc]{IEEEtran}
\else
\documentclass[11pt]{article}
\advance \topmargin by -\headheight
\advance \topmargin by -\headsep
\evensidemargin \oddsidemargin
\newcommand{\IEEEcompsocitemizethanks}{\thanks}
\newcommand{\IEEEmembership}[1]{\relax}
\newcommand{\IEEEcompsocthanksitem}{\relax} 
\newenvironment{IEEEproof}{\noindent{\bf Proof:}}{\hspace*{\fill}\rule{6pt}{6pt}\bigskip}
\fi
\usepackage{graphicx}
\usepackage{paralist}
\usepackage{amssymb}
\usepackage{url}
\usepackage[noend]{algorithmic}
\usepackage{cite}

\newtheorem{theorem}{Theorem}

\newtheorem{lemma}{Lemma}

\makeatletter
\def\@begintheorem#1#2{\sl \trivlist \item[\hskip \labelsep{\bf #1\ #2:}]}
\def\@opargbegintheorem#1#2#3{\sl \trivlist
      \item[\hskip \labelsep{\bf #1\ #2\ #3:}]}
\makeatother

\newcommand{\Insert}{\textbf{{Insert}}}
\newcommand{\Delete}{\textbf{{Delete}}}

\newcommand{\LS}{\textbf{{ListStragglers}}}

\begin{document}
\title{Straggler Identification in Round-Trip \\
      Data Streams via Newton's Identities and \\
      Invertible Bloom Filters}
\author{David~Eppstein
       and
       Michael~T.~Goodrich~\IEEEmembership{Fellow,~IEEE}%
\IEEEcompsocitemizethanks{\IEEEcompsocthanksitem David~Eppstein
and Michael T.~Goodrich are with the Department
of Computer Science, Univesity of California, Irvine, CA 92697-3435.
\hfil\break
\IEEEcompsocthanksitem 
E-mail and web pages: see \texttt{http://www.ics.uci.edu/\char`\~eppstein}
and \texttt{http://www.ics.uci.edu/\char`\~goodrich}.%
}}
\date{}

\maketitle
\begin{abstract}

In this paper,
we study the \emph{straggler identification} problem, in which an
algorithm must determine the identities of the remaining members of a
set after it has had a large number of insertion and deletion
operations performed on it, and now has relatively few remaining
members.  
\ifFull
The goal is to do this in $o(n)$ space, where $n$ is the
total number of identities.  Straggler identification has applications,
for example, in determining the unacknowledged packets in a
high-bandwidth multicast data stream.  We provide a deterministic
solution to the straggler identification problem that uses only
$O(d\log n)$ bits, based on a novel application of Newton's identities
for symmetric polynomials.  This solution can identify any subset of
$d$ stragglers from a set of $n$ $O(\log n)$-bit identifiers, assuming
that there are no false deletions of identities not already in the
set.  Indeed, we give a lower bound argument that shows that any
small-space deterministic solution to the straggler identification
problem cannot be guaranteed to handle false deletions.  Nevertheless,
we provide a simple randomized solution using $O(d\log n\log
(1/\epsilon))$ bits that can maintain a multiset and solve the
straggler identification problem, tolerating false deletions, where
$\epsilon>0$ is a user-defined parameter bounding the probability of an
incorrect response.  
This randomized solution is based on a new type of
Bloom filter, which we call the \emph{invertible Bloom filter}.

\fi
\end{abstract}

\section{Introduction}
Imagine a security guard, who we'll call Bob, working at a 
large office building.
Every day, Bob comes to work before anyone else, 
unlocks the front doors, and then staffs the front desk.
After unlocking the building,
Bob's job is to check in each of a set of $n$ workers when he or she enters the
building and check each worker out again when he or she leaves.
Most workers leave the building by 6pm, when Bob's shift ends.
But, at the end of Bob's shift, there may be a small number, at most $d<<n$, of
\emph{stragglers}, who linger in the building working overtime.
Before Bob can leave for home,
he must tell the night guard the ID numbers of all the stragglers.
The challenge is that Bob has only a small clipboard of size $o(n)$
to use as a ``scratch space'' for recording information as
workers come and go.
That is, Bob does not have enough room on his clipboard to write
down all the ID numbers of the workers as they arrive and 
to check off these numbers again as they leave.
Of course, he also has to deal with the fact that some of the $n$
workers may not come to work at all on any given day.
The question we address in this paper is, 
``What information can Bob, the security guard, record as he
checks workers in and out so that he may
identify all the stragglers at the end of his shift,
using a scratch space of size only $o(n)$?''

Formally, suppose that we are
given a universe $U=\{x_1,x_2,\ldots,x_n\}$ of 
unique, positive identifiers, each representable with $O(\log n)$ bits.
Given an upper bound parameter $d<<n$, 
the \emph{straggler identification problem} is the problem of
designing an indexing structure for 
a database that uses $o(n)$ bits and efficiently 
supports the following operations on 
a dynamic and initially-empty subset $S$ of~$U$:
\begin{itemize}
\item {\Insert} $x_i$: Add the identifier $x_i$ to $S$. Prior to the update, $x_i$ should not belong to $S$; the effect of the insert operation is undefined if $x_i\in S$.
\item {\Delete} $x_i$: Remove the identifier $x_i$ from $S$. Prior to the update, $x_i$ should belong to $S$; the effect of the delete operation is undefined if $x_i\notin S$.
\item {\LS}: Test whether $|S|\le d$, and if so, list all the elements of $S$.
\end{itemize}
A solution to the straggler identification can be used to list the contents of $S$ when $|S|\le d$, but makes no such guarantees when $|S|>d$. In our solutions to this problem we will assume, without loss of generality,
that $d$ is small enough that $d\log (n/d)$ is $o(n)$.
If, on the contrary, $d$ is larger, then the problem is not solvable in $o(n)$ bits, since we need to store $\Omega(d\log (n/d))$ bits in order to distinguish among the different possible valid answers to a {\LS} query. Moreover, if $d$ is close to $n$ we might as well just
store all the elements of $S$ explicitly using a single bit per element.
However, by requiring that $d$ be small and that our structure use $o(n)$ bits of memory, we focus our attention on implicit representations of $S$.

In addition to our motivating example of Bob, the security guard (which
also applies to other in-and-out physical environments, like amusement parks),
the straggler identification problem has the following potential information-processing
applications:
\begin{itemize}
\item
In a high bandwidth data stream, a server sends
packets to many different clients, which send
acknowledgments back to the server identifying each packet that was
successfully received. 
The server then needs to identify and re-send the packets to clients that did 
not successfully receive them.
This round-trip data stream
application is an instance of the straggler identification problem,
since we expect most of the packets to be sent successfully and we
would like to minimize the space needed per client at the server for
unacknowledged packet identification.
\item
In heterogeneous Grid computations, a supervisor sends
independent tasks out to Grid participants, who, under normal
conditions, perform these tasks and return the results to the
supervisor. There may be a few participants, however, who crash,
are disconnected 
from the network, or otherwise fail to perform their tasks.
The supervisor would like to identity
the tasks without responses, so that they can be sent to other
participants for completion.
\item
At the beginning of the school year in a public grade school,
teachers distribute textbooks to students. At the end of the year, most students return those books. But there may be a
few stragglers who do not return their textbooks, and
the teacher would, with low computational
overhead, like to identify those students.
\item
A software company issues pseudo-random serial numbers to users who 
who download their software, with an implied commitment to 
return payment within a week. Most of these users do indeed return 
such a payment, tagged with their serial numbers. 
But a few do not, and we would like to identify the serial numbers of the
users who have not returned payment.
\end{itemize}

Given these motivating applications, the goal of 
the straggler identification problem is to design a database
indexing scheme that uses as few bits as possible, with
reasonable running times for performing the {\Insert}, {\Delete},
and {\LS} operations.

\subsection{New Results}
In this paper, we study the straggler identification problem, showing
that it can be solved with small space and fast update times.
We provide the following results:
\begin{itemize}
\item In Section~\ref{sec:polys}, We describe a deterministic solution to the straggler identification problem, which uses $O(d\log n)$ bits to
represent the dynamic set $S$ of $O(\log n)$-bit identifiers.
Our solution is based on a novel application of Newton's identities
and allows for insertions and deletions to be performed in 
$O(d\log^{O(1)} n)$ time. It allows the {\LS} operation to be performed in
time polynomial in $d$ and $\log n$.
This solution does not allow (false) {\Delete} $x$ operations
that have no matching {\Insert} $x$ operations, however: our algorithm does not detect false deletions, and may produce unpredictable results if it is asked to handle an update sequence in which false deletions occur.
\item As a partial explanation of our inability to handle false deletions, we prove in Section~\ref{sec:negative} a lower bound showing that no
deterministic algorithm for the straggler detection problem with sublinear space can guarantee correctness in scenarios allowing false deletions. Thus, this drawback of our algorithm should come as no surprise.
\item Despite this impossibility result, we provide a second solution to the straggler
identification problem, in Section~\ref{sec:bloom}. Our solution is a simple randomized algorithm that uses 
$O(d\log n\log (1/\epsilon))$ bits and tolerates false deletions,
where $\epsilon>0$ is a user-defined error probability bound. Our algorithm can handle any sequence of updates, and has probability at most $\epsilon$ of being unable to correctly answer a {\LS} query.
This solution is based on a novel extension to the 
counting Bloom filter~\cite{bmpsv-iccbf-06,fcab-scswa-00},
which itself is a dynamic,
cardinality-based extension to 
the well-known Bloom filter data structure~\cite{B70}
(see also~\cite{bm-nabfs-05}).
We refer to our extension as the \emph{invertible Bloom filter},
because, unlike the standard Bloom filter 
and its counting extension---which provide 
a degree of data privacy protection---the invertible
Bloom filter allows for the efficient enumeration of its contents if
the number of items it stores is not too large.
This might seem like a violation of the spirit of a Bloom
filter, which was invented specifically to avoid the space
needed for content enumeration.
Nevertheless, the invertible Bloom filter is useful for straggler
identification, because it can at one time
represent, with small space, a multiset that is too large to
enumerate, and later, after a series of deletions have been performed, 
provide for the efficient listing of the remaining elements.
\end{itemize}

\subsection{Related Work}
Our work is most closely related to the ``deterministic $k$-set
structure'' of Ganguly and Majumder~\cite{gm-pods-06,Ganguly200827}.  This
structure solves the straggler detection problem, and unlike our
solution it allows items to have multiplicity greater than one. This
solution, like our deterministic algorithm, disallows false deletions
and is based on the arithmetic of finite fields. However the most
space-efficient version of their solution uses roughly twice as many
bits as ours, and their decoding times are slower: ignoring logarithmic
factors, their structure's {\LS} queries take $O(d^3)$ or $O(d^4)$
time, compared to $O(d^2)$ for ours. An additional technical difference
is that, for the algorithm of Ganguly and Majumder, the parameter $k$
(analogous to our $d$) measures the number of distinct stragglers,
while for us it measures the total number of stragglers. Independently
of our work, Ganguly and Majumder added to the journal
version of their paper a lower bound similar to ours proving the
impossibility of straggler detection with false 
deletions~\cite{Ganguly200827}.

Our deterministic solution is also related to work on set
reconciliation in communication complexity~\cite{mtz-itit-03}. The set
reconciliation problem is the problem of finding the union of two
similar sets, held by two different communicating parties, with an
amount of communication close to the size of the symmetric difference
of the two sets. A solution to the straggler detection problem that
allows false deletions could be used to solve the set reconciliation
problem, as follows: the first party inserts all of the elements of its
set into a straggler detection data structure and then communicates the
structure to the second party, who deletes all of the elements of its
set. The remaining small numbers of stragglers and false deletions
represent the symmetric difference of the two sets. However, Minsky et
al.~\cite{mtz-itit-03} present a protocol for the set reconciliation
problem that is more closely related to our deterministic straggler
detection algorithm (which does not allow false deletions) than t

Some additional existing work
can be adapted to solve the straggler identification problem.
For example, Cormode and Muthukrishnan~\cite{cm-whwn-05} study 
the problem of
identifying the $d$ highest-cardinality members of a dynamic multiset.
Their solution can be applied to the straggler
identification problem, since whenever there are $d$ 
or fewer elements in the set, then all elements are of relatively 
high cardinality. Their result is a randomized data structure that
uses $O(d\log^2 n\log (1/\epsilon))$ bits to perform updates in 
$O(\log^2 n\log (1/\epsilon))$ time and can be adapted to answer
{\LS} queries in $O(d\log^2 n\log (1/\epsilon))$ time 
(in terms of their bit complexities),
where $\epsilon>0$ is a user-defined parameter bounding the
probability of a wrong answer.

Also relevant is prior work on combinatorial group testing (CGT),
e.g., see~\cite{colbourn99applications,dgv-gfsao-03,%
dh-cgtia-00,dh-pdngt-06,egh-icgtr-05,fkkm-gtpse-97,gh-epads-06,HS87},
and multiple access channels (MAC),
e.g., see~\cite{c-tapbc-79,gp-crpra-82,gl-emcma-83,gw-lbtnw-85,h-sacdc-84,%
p-bppma-81,rv-ccac-94,t-rckm-80}.
In combinatorial group testing, there are $d$
``defective'' items in a set $U$ of $n$ objects, for which
we are allowed to perform \emph{tests}, which involve forming a
subset $T\subseteq U$ and asking if there are any defective items in $T$.
In the standard combinatorial group testing problem, the outcome is
binary---either $T$ contains defective items or it does not.
The objective is to identify all $d$ defective items.
The combinatorial group testing algorithms that are most relevant to straggler
identification are \emph{nonadaptive}, in that they must ask all of their tests,
$T_1,T_2,\ldots, T_m$, in advance.
Such an algorithm can be converted to solve
the straggler identification problem by creating a counter $t_i$
for each test $T_i$.
On an insertion of $x$, we would increment each $t_i$ such that
$x\in T_i$.
Likewise, on a deletion of $x$, we would decrement each $t_i$ such
that $x\in T_i$.
The tests with non-zero counters would be exactly
those containing our objects of interest, and the nonadaptive
combinatorial group testing algorithm could then be used to identify them.
Unfortunately, these algorithms don't translate into efficient
straggler-identification methods,
as the best known nonadaptive combinatorial group testing algorithms 
(e.g., see~\cite{dh-cgtia-00,dh-pdngt-06})
use $O(d^2 \log n)$ tests, which would
translate into a straggler solution needing $O(d^2\log^2 n)$ bits.

The multiple access channel problem is similar to the combinatorial group testing problem, except that the
items of interest are no longer ``defective''---they are $d$ devices,
out of a set $U$, wishing to broadcast a message on a common channel.
In this case a ``test'' is a time slice where members
of a subset $T\subseteq U$ can broadcast.
Such an event has a three-way outcome, in that there can be $0$
devices that use this time slice, $1$ device that uses it (in which
case it is identified and taken out of the set of potential broadcasters), 
or there can be $2$ or
more who attempt to use the channel, in which case none succeed (but
all the potential broadcasters learn that $T$ contains at least two
broadcasters).
Unfortunately, traditional multiple access channel algorithms are adaptive, so do not
immediately translate into straggler identification
algorithms.

Nevertheless, we can extend the multiple access channel approach 
further~\cite{gp-crpra-82,p-bppma-81,rv-ccac-94,t-rckm-80},
so that each test $T$ returns the actual number of items of 
interest that are in $T$. 
This extension gives rise to a
\emph{quantitative} version of combinatorial group testing (e.g., see~\cite{dh-cgtia-00}, Sec.~10.5).
Unfortunately, previous approaches to the quantitative combinatorial group testing problem are either
non-constructive~\cite{p-bppma-81},  
adaptive~\cite{gp-crpra-82,p-bppma-81,rv-ccac-94,t-rckm-80}, or
limited to small values of $d$.
We know of no nonadaptive quantitative combinatorial group testing algorithms 
for $d\ge 3$, and the ones for $d=2$ don't translate into efficient
solutions to the straggler identification problem (e.g.,
see~\cite{dh-cgtia-00}, Sec.~11.2).

\section{Straggler Detection via Symmetric Polynomials}
\label{sec:polys}

We now describe a deterministic algorithm for straggler detection using near-optimal memory. The algorithm is algebraic in nature: it stores as its snapshot of the data stream a collection of \emph{power sums}. The decoding algorithm for this information uses Newton's identities to convert these power sums into the coefficients of a polynomial that has the stragglers as its roots, and finds the roots of this polynomial. In order to control the time complexity of the root-finding algorithm used as a subroutine in our {\LS} operations and the space complexity for storing the power sums, we perform our operations in a carefully chosen finite field $GF[p^e]$.

As a notational simplification, we use $\tilde O(x)$ as a shorthand for $O(x\log^{O(1)} x)$. Using this notation, we ignore terms in our running times that are logarithmic in the overall time bound.

\subsection{Newton's Identities}

A \emph{symmetric polynomial} in a set $S$ of variables $\{x_1,x_2,\ldots\}$ is a multivariate polynomial that maintains the same overall value whenever the values of the variables in $S$ are permuted arbitrarily. Two particularly important families of symmetric polynomials are the \emph{elementary symmetric polynomials} $\sigma_k$, the sums of all $k$-tuples of distinct variables
$$\sigma_1=x_1+x_2+x_3+\ldots,$$
$$\sigma_2=x_1x_2 + x_1x_3 + x_2x_3+\ldots,$$
$$\sigma_3=x_1x_2x_3 + x_1x_2x_4 + x_1x_3x_4+\ldots,$$
$$\vdots$$
and the \emph{power sums} $s_k=\sum x_i^k$:
$$s_1=x_1+x_2+x_3+\ldots,$$
$$s_2=x_1^2+x_2^2+x_3^2+\ldots,$$
$$s_2=x_1^3+x_2^3+x_3^3+\ldots,$$
$$\vdots$$
The significance of these polynomials for straggler detection is that the power sums may be maintained easily by a streaming algorithm, whereas the elementary symmetric polynomials may be combined to form the coefficients of a univariate polynomial that has the stragglers as its roots.

\emph{Newton's identities} (e.g. see~\cite{cls-iva-92}) provide a formula for computing the power sums from the elementary symmetric polynomials:
$$s_k-k(-1)^k\sigma_k=-\sum_{i=1}^{k-1} (-1)^i \sigma_i s_{k-i}.$$
That is,
\begin{eqnarray*}
s_1-\ \,\sigma_1  &=& 0 \\
s_2+2\sigma_2 &=& \sigma_1s_1 \\
s_3-3\sigma_3 &=& \sigma_1s_2-\sigma_2s_1 \\
s_4+4\sigma_4 &=& \sigma_1s_3-\sigma_2s_2+\sigma_3s_1 \\
s_5-5\sigma_5 &=& \sigma_1s_4-\sigma_2s_3+\sigma_3s_2-\sigma_4s_1,
\end{eqnarray*}
and so on. These equations hold over any field.

In our application, we need to invert this system of equations, computing the value of the elementary symmetric polynomials from the power sums. In the presentation of the identities above, each equation is a linear combination of the elementary symmetric polynomial of order $k$, the power sum of order $k$, and terms computed from symmetric polynomials of both types of lower order. Therefore, we may use these identities to compute the elementary symmetric polynomials $\sigma_k$ from the power sums, in order by $k$, by rearranging the equations so that the left hand side is the symmetric polynomial $\sigma_k$ and the right hand side is $1/k$ times a linear combination of known and previously computed terms.
However, this rearranged system of identities is no longer valid over all fields: computing $\sigma_k$ from the identities above requires a division by the integer $k$, so if we are to perform our computations within a finite field $GF[p^e]$ then $k$ must not be divisible by the order $p$ of the field.

\subsection{Arithmetic in Finite Fields}

For the correctness of our straggler detection algorithm, we are free to perform our arithmetic operations within any finite field in which the order of the field is large enough to allow Newton's identities to be inverted; however, different choices of field will lead to different running times for the root-finding subroutine in our algorithm for handling {\LS} queries. Thus, rather than working in the integers modulo a prime $p$ that is larger than our universe size $n$, it will turn out to be more efficient to work in a finite field $GF[p^e]$ of a smaller order $p$. We briefly summarize the necessary facts about computational arithmetic in such fields; for a more detailed explanation, see e.g.~\cite{Coh-93}.

As is standard for this sort of computation, we represent each value $x$ in $GF[p^e]$ as a univariate polynomials of degree at most $e-1$ in a variable $\theta$, with coefficients that are integers modulo~$p$; that is,
$$x=x_0+x_1\theta+x_2\theta^2+\cdots+x_{e-1}\theta^{e-1},$$
where each coefficient $x_i$ is an integer modulo~$p$. Therefore, values in the field $GF[p^e]$ may be represented using $e\lceil\log_2 p\rceil$ bits per value.
These polynomials are taken modulo a \emph{monic irreducible polynomial} $$Z(\theta)=Z_0+Z_1\theta+Z_2\theta^2+\cdots+Z_{e-1}\theta^{e-1}+\theta^e.$$
This modulus~$Z$ may be found e.g. by a deterministic algorithm of Shoup~\cite{Sho-MC-90}.
The sum or difference of any two polynomials representing values in $GF[p^e]$ may be computed by coordinatewise modulo-$p$ addition:
$$x+y=(x_0+y_0)+(x_1+y_1)\theta+(x_2+y_2)\theta^2+\cdots.$$

To multiply two values in $GF[p^e]$, one may use a convolution-based polynomial multiplication algorithm to produce a single product polynomial of degree $2(e-1)$, and then reduce the product modulo $Z$.
Working modulo $Z$ is equivalent to constraining $\theta$ to satisfy the equation $Z(\theta)=0$, that is,
$$\theta^e=-(Z_0+Z_1\theta+Z_2\theta^2+\cdots+Z_{e-1}\theta^{e-1}).$$
This equation allows the product polynomial, of degree $2(e-1)$, to be reduced to a polynomial of degree at most $e-1$ in a sequence of $O(\log e)$ steps. In the $i$th-from-last reduction step we split the reduced polynomial $q_i(\theta)$
into two parts:
$$q_i(\theta)=r_i(\theta)+\theta^{e-1+2^i} h_i(\theta),$$
where $h_i$ has degree $2^i$ and $r_i$ has degree $e-2+2^i$; this split may be accomplished simply by partitioning the coefficients of $q_i$ according to their degrees. We then compute the product of $h_i$ with a polynomial of degree $e-1$ equal in value (modulo $Z$) to $\theta^{e-1+2^i}$, and replace $q_i$ with a polynomial $q_{i-1}$, the sum of this product with $r_i$.
In this way, multiplication in $GF[p^e]$ may be accomplished using $O(\log e)$ calls to a polynomial multiplication subroutine.
A modified version of the Sch\"onhage--Strassen integer multiplication algorithm allows each of these calls to be accomplished in $\tilde O(e)$ modulo-$p$ operations~\cite{CanKal-AI-91,Nus-ASSP-80,SchStr-C-71}.

We do not need to perform divisions by arbitrary values in $GF[p^e]$, but
our algorithms do involve division of values in $GF[p^e]$ by integers in the range $[2,p-1]$; this may be done by dividing each coefficient of the value independently by the given integer, modulo~$p$.

Therefore, each field operation may be performed in bit complexity $\tilde O(e\log p)$.

\subsection{The Algorithm}

\begin{theorem}
There is a deterministic streaming straggler detection algorithm using $(1+o(1))(d+1)\log n$ bits of storage, such that {\Insert} and {\Delete} operations can be performed in bit complexity $\tilde O(d\log n)$, and such that {\LS} operations can be performed in bit complexity $\tilde O(d\log^3 n + d^2\log n + d^{3/2}\log^2 n\min(d,\log n))$.
\end{theorem}

\begin{IEEEproof}
We let $p$ be a prime number, larger than $d$ but at most $O(d)$, and let $e=\lceil \log_p (n+1)\rceil$ so that
$p^e>n$.  We perform all operations of the algorithm in the field
$GF[p^e]$, and interpret all identifiers in the straggler detection
problem as values in this field. The number of bits needed to
represent a single value in $GF[p^e]$ is $(1+o(1))\log_2 n$, and, with this
choice of $p$ and $e$, each arithmetic operation in the field may be
performed in bit complexity $\tilde O(\log n)$.

Define the power sums
$$s_k(S) = \sum_{x_i\in S} x_i^k$$
(where $x_i$ and $s_k$ belong to $GF[p^e]$, except for $s_0$ which we store as a $\log n$ bit integer).
Our streaming algorithm stores $s_k(S)$ for $0\le k\le d$. As $s_0(S)$ is the number of stragglers, we can easily compare the number of stragglers to~$d$.

To update the power sums after an insertion of a value $x_i$, we simply add $x_i^k$ to each power sum $s_k$; this requires $O(d)$ arithmetic operations in $GF[p^e]$ to compute the powers of $x_i$ and perform the additions. Similarly, to delete $x_i$, we subtract $x_i^k$ from each power sum $s_k$.

At any point in the algorithm, we may define a polynomial in $GF[p^e][x]$,
$$P(x)=\prod_{x_i\in S}(x-x_i)=\sum_{k=0}^{|S|}(-1)^k\sigma_kx^{|S|-k},$$
where $\sigma_k$ is the $k$th elementary symmetric function of $S$. By using Newton's identities, we may calculate the coefficients of $P$ in sequence from the power sums and the earlier coefficients, using $O(d^2)$ arithmetic operations to compute all coefficients. Thus, this stage of the {\LS} operation takes bit complexity $\tilde O(d^2\log n)$.

Finally, to determine the list of stragglers, we find the roots of the polynomial $P(x)$ that has been determined as above. The deterministic root-finding algorithm of Shoup~\cite{s-fdafpff-91} solves this problem in $\tilde O(d\log^2 n + d^{3/2}\log n\min(d,\log n))$ field operations; multiplying this by the $\tilde O(\log n)$ bound on the number of bit operations per field operation gives the $\tilde O(d\log^3 n)$ and $\tilde O(d^{5/2}\log n\min(d,\log n))$ terms in the statement of the theorem.
Thus, the overall bit complexity bound is as stated.
\end{IEEEproof}

We note that a factor of $d^{1/2}$ in Shoup's algorithm~\cite{s-fdafpff-91}
occurs only when $p$ has an unexpectedly long repeated subsequence in its sequence of quadratic characters. Per the discussion in Shoup's paper, it seems likely that a more careful choice of $p$ can eliminate this factor, simplifying the time bound for the {\LS} operation to $\tilde O(d\log^3 n+d^2\log n)$. If this is possible, it would be an improvement when $d$ lies in the range of values from $\log^{2/3} n$ to $\log^2 n$.

For $d=2$, the root finding algorithm may be replaced by the quadratic formula for solving a degree-two polynomial, and similarly for $d\le 4$ the root finding algorithm may be replaced by the closed-form formulas for the solutions of cubic and quartic polynomials.

\section{Impossibility Results in the Presence of False Deletions}
\label{sec:negative}

So far, we have assumed that an element deletion can occur only if a corresponding insertion has already occurred. That is, the only anomalous data patterns that might occur are insertions that are not followed by a subsequent deletion. What can we say about more general update sequences in which insertion-deletion pairs may occur out of order, multiple times, or with a deletion that does not match an insertion? We would like to have a streaming data structure that handles these more general event streams and allows us to detect small numbers of anomalies in our insertion-deletion sequences.

Formally, define a \emph{signed multiset} over a set $S$ to be a map $f$ from $S$ to the integers, where $f(x)$ is the number of occurrences of $x$ in the multiset. To insert $x$ into a signed multiset, increase $f(x)$ by one, while to delete $x$, decrease $f(x)$ by one. Thus, any sequence of insertions and deletions, no matter how ordered, produces a well-defined signed multiset. We wish to find a streaming algorithm that can determine whether all but a small number of elements in the signed multiset have nonzero values of $f(x)$ and identify those elements. But, as we show, for a natural and general class of streaming algorithms, even if restricted to signed multisets in which each $x$ has $f(x)\in\{-1,0,1\}$, we cannot distinguish the empty multiset (in which all $f(x)$ are zero) from some nonempty multiset. Therefore, it is impossible for a deterministic streaming algorithm to determine whether a multiset has few nonzeros.

The signed multisets form a commutative group, isomorphic to ${\mathbb Z}^{|S|}$, which we will represent using additive notation: $(f+g)(x) = f(x)+g(x)$. Call this group $M$. Define a \emph{unit multiset} to be a signed multiset in which all values $f(x)$ are in $\{-1,0,1\}$; the unit multisets form a subset of $M$, but not a subgroup.

Suppose a streaming algorithm maintains information about a signed multiset, subject to insertion and deletion operations. We say that the algorithm is \emph{uniquely represented} if the state of the algorithm at any time depends only on the multiset at that time and not on the ordering of the insertions and deletions by which the multiset was created. That is, there must exist a map $u$ from $M$ to states of the algorithm. Intuitively, this is a natural requirement on an efficient streaming algorithm, because the additional bits required to allow the representation of multiple different states for the same multiset represent wasted storage space. The deterministic straggler detection algorithm of the previous section, for instance, is uniquely represented.

Define a binary operation $+$ on states of a uniquely represented multiset streaming algorithm, as follows. If $a$ and $b$ are states, let $A$ and $B$ be signed multisets such that $u(A)=a$ and $u(B)=b$, and let $a+b=u(A+B)$. 

\begin{lemma}
\label{lem:representivity}
If a streaming algorithm is uniquely represented, and $u(P)=u(Q)$, then $u(P+R)=u(Q+R)$.
\end{lemma}

\begin{IEEEproof}
Let $s$ be a sequence of updates that forms $R$. Then $s$ transforms $u(P)$ to $u(P+R)$ and $u(Q)$ to $U(Q+R)$. Since $u(P)=u(Q)$, $u(P+R)$ and $u(Q+R)$ result from applying the same sequence of updates to the same initial state, and therefore must equal each other.
\end{IEEEproof}

\begin{lemma}
\label{lem:state-group}
The addition operation on states defined above is well-defined independently of how the representative multisets $A$ and $B$ are chosen, the states of the streaming algorithm form a commutative group under this operation, and u is a group homomorphism.
\end{lemma}

\begin{IEEEproof}
Independence from the choice of representation 
is Lemma~\ref{lem:representivity}: 
if $A$ and $A'$ represent the same state, and $B$ and $B'$ represent the same state, then by two applications of 
Lemma~\ref{lem:representivity} we may substitite $A$ for $A'$ and $B$ for $B'$, showing that $A+B$ and $A'+B'$ represent the same state.

Associativity and commutativity follow from the associativity and commutativity of the corresponding group operation on $M$: if two states are represented by the elements $A$ and $B$ of $M$, then the sum of the two states (in either order of summation) is represented by $A+B=B+A$, where the equality is just commutativity within $M$. Similarly, if three states are represented by the elements $A$, $B$, and $C$ of $M$, then the sum of the three states (in either of two ways of grouping the sum) is represented by $(A+B)+C=A+(B+C)$, where again the equality is just commutativity within $M$.

By Lemma~\ref{lem:representivity}, $u(A)+u(-A)=u(0)$ and $u(A)+u(0)=u(A)$, so $u(0)$ satisfies the axioms of a group identity.

Because addition of states satisfies associativity, commutativity, and identity, we have defined a commutative group. That $u$ is a homomorphism follows from the way we have defined our group operations as the images by $u$ of group operations in $M$.
\end{IEEEproof}

\begin{theorem}
Any uniquely represented multiset streaming algorithm for a multiset on $n$ items, with fewer than $n$ bits of storage, will be unable to distinguish between the empty set and some nonempty unit multiset.
\end{theorem}

\begin{IEEEproof}
Suppose there are $k<n$ bits of storage, so that the data structure has at most $2^k$ possible states. By the pigeonhole principle, two different sets $A$ and $B$, when interpreted as multisets and mapped to states, map to the same state $u(A)=u(B)$. Then by Lemma~\ref{lem:state-group}, $u(A-B)=u(\emptyset)$. $A-B$ is a nonempty unit multiset that cannot be distinguished from the empty set.
\end{IEEEproof}

By applying similar ideas, we can prove a similar impossibility result without making our unique representativity assumption about the nature of the streaming algorithm.

\begin{theorem}
No deterministic streaming algorithm with fewer than $n$ bits of storage can distinguish a stream of matched pairs of insert and delete operations over a set of $n$ items from a stream of insert and delete operations that are not matched in pairs.
\end{theorem}

\begin{IEEEproof}
Suppose that we have a deterministic streaming data structure with $k<n$ bits of storage. For any set $A$, let $f(A)$ denote the state of the data structure on a stream that starts with an empty set and inserts the items in $A$ in some canonical order. By the pigeonhole principle there exist two sets $A$ and $B$ such that $A\ne B$ but such that $f(A)=f(B)$. Let $s_{PQ}$ ($P,Q\in\{A,B\}$) be the operation stream formed by inserting the items in set $P$ followed by deleting the items in set $Q$. Then the streaming algorithm must have the same state after stream $s_{AA}$ as it does after stream $s_{BA}$, but $s_{AA}$ consists of matched insert-delete pairs while $s_{BA}$ does not.
\end{IEEEproof}

\ifFull
Another way of stating this result is that, for any deterministic
streaming algorithm, some nonempty set $A$ must be indistinguishable
from the empty set, so it is impossible to always correctly answer
queries that should give different answers for empty and nonempty sets.
This argument doesn't apply to
a randomized streaming algorithm, however, as it may be very unlikely
that any particular set queried by the algorithm has this property of
being indistinguishable from empty. 
This observation motivates the results in the
following section, in which we describe streaming algorithms for a
multiset version of the straggler detection problem that use randomness
to evade the limitations of our impossibility results. As with previous
randomized streaming algorithms, our algorithm may give mistaken
answers to queries, but it is highly unlikely that any particular query
is answered incorrectly.
\fi

\newcommand{\Bit}{\texttt{bit}}
\newcommand{\Count}{\texttt{count}}
\newcommand{\Idsum}{\texttt{idSum}}
\newcommand{\Hashsum}{\texttt{hashSum}}
\section{Invertible Bloom Filters}
\label{sec:bloom}
\begin{figure*}[t]
\centering\includegraphics[scale=0.9]{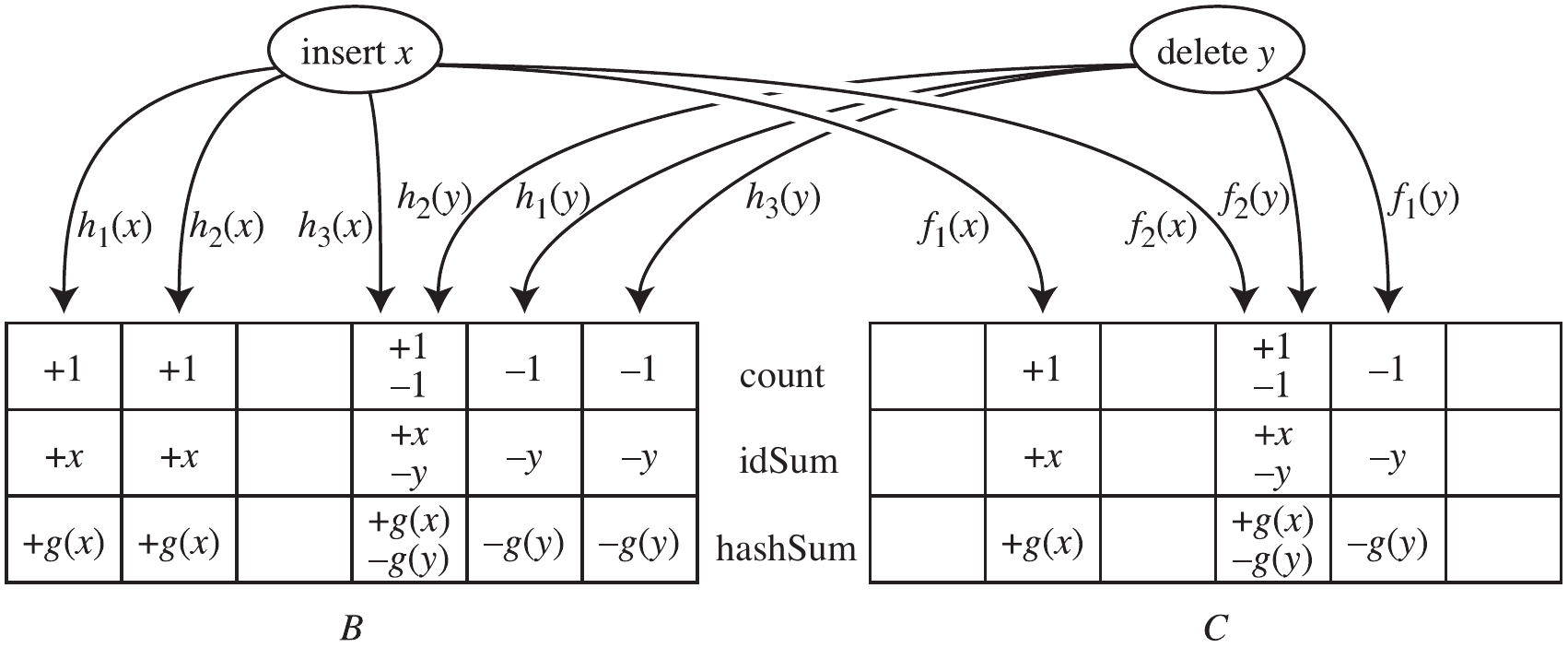}
\caption{The updates performed by insertion and deletion operations in an invertible Bloom filter.}
\label{fig:indel}
\end{figure*}
The standard Bloom filter~\cite{B70} is a randomized data structure
for approximately representing a set $S$ subject to insertion 
operations and membership queries.

Given a parameter $d$ on the expected size of $S$ and an error
parameter $\epsilon>0$, a standard Bloom filter
consists of a hash table $B$ containing $m=O(d\log (1/\epsilon))$ 
single-bit cells (which we denote as a ``\Bit'' field), together with
$k=\Theta(\log (1/\epsilon))$ random hash functions $\{h_1,\ldots,h_k\}$
that map elements of $S$ to integers in the range $[0,m-1]$.
 
Initially each cell contains the value $0$.
An insertion of an element $x$ into the standard Bloom filter is performed by setting each 
$B[h_i(x)].\Bit$ to $1$, for $i=1,\ldots,k$.
Likewise, testing for membership of $x$ in $S$ amounts to testing that 
there is no $i\in\{1,\ldots,k\}$ such that $B[h_i(x)].\Bit = 0$.
If one sets the constant factor in the formulas for $m$ and $k$ appropriately, one can cause the
probability that this data structure returns a false positive to any single membership query
(that is, that any particular element not in $S$ is erroneously identified as belonging to $S$) 
to become less than the error parameter $\epsilon$ (e.g., see~\cite{bgkmmmst-otfpr-07}).

Standard Bloom filters do not allow elements, once inserted, to be deleted from~$S$. To remedy this inability,
the counting Bloom filter~\cite{bmpsv-iccbf-06,fcab-scswa-00}
extends the standard Bloom filter by replacing each ``{\Bit}'' 
cell of $B$ with a counter cell, ``{\Count}''
(as before, initialized to $0$ for each cell).
An insertion of item $x$ is performed by incrementing each 
$B[h_i(x)].\Count$ by $1$, for $i=1,\ldots,k$.
Such a structure also supports the 
deletion of an item $x$, by decrementing each cell
$B[h_i(x)].\Count$ by $1$, for $i=1,\ldots,k$.
Answering a membership query is similar to that for the standard Bloom filter,
and is performed by testing that 
there is no $i\in\{1,\ldots,k\}$ such that $B[h_i(x)].\Count = 0$. The error analysis is the same as for standard Bloom filters. However, although counting Bloom filters can be used to map any set to a fully dynamic membership testing data structure, the map cannot be inverted efficiently: it is not obvious how to find the members of a set represented by a counting Bloom filter other than by testing membership for all elements in the universe.

\subsection{The Indexing Scheme for the Invertible Bloom Filter}

The \emph{invertible Bloom filter} extends the counting Bloom filter,
in several ways, and allows us to solve the straggler identification
problem even in the presence of false deletions.
It requires that we use three additional random hash
functions, $f_1$, $f_2$, and $g$, in addition to the $k$ hash
functions, $h_1,\ldots,h_k$, used for $B$ above.
The functions, $f_1$ and $f_2$ map integers in $[0,n]$ to integers in
$[0,m]$. The function $g$ maps integers in $[0,n]$ to integers in 
$[0,n^2]$.
In addition, we add two more fields to each Bloom filter cell, $B[i]$:
\begin{itemize}
\item
An ``{\Idsum}'' field, which stores the sum of all the 
elements, $x$ in $S$, for $x$'s that map to the cell $B[i]$.
Note that if $B[i]$ stores $m$ copies of a value $x$ (and no other
values), then $B[i].\Idsum = mx$.
\item
A ``{\Hashsum}'' field, which stores the sum of all the 
hash values, $g(x)$, for $x$'s that map to the cell $B[i]$.
Note that if $B[i]$ stores $m$ copies of a value $x$ (and no other
values), then $B[i].\Hashsum = mg(x)$.
\end{itemize}
The {\Idsum} field must be of size at least $\log n + \log d$ bits, so that
it can store $d$ ID's and the {\Hashsum} field should be 
of size at least $2\log n + \log d$ bits, so that it can
store $d$ numbers in the range $[0,n^2]$.
We allow these fields to overflow, in the case that there are
more than $d$ numbers summed in either field.
But we require that addition and subtraction remain inverses of each
other, so that it is always the case that $(a+b)-b=a$
and $(a-b)+b=a$.

In addition to these fields in $B$, 
we create a second Bloom filter, $C$, which has the
same number of ({\Count}, {\Idsum}, and {\Hashsum}) fields as $B$, but uses only
the functions $f_1$ and $f_2$ to map elements of $S$ to its cells.
That is, $C$ is a secondary augmented counting Bloom filter with the
same number of cells as $B$, but with only two random hash functions,
$f_1$ and $f_2$, to use for mapping purposes.
Intuitively, $C$ will serve as a fallback Bloom filter for
``catching'' elements that are difficult to recover using $B$ alone.
Finally, in addition to these fields, we maintain a global {\Count}
variable, initially $0$.
Each of our {\Count} fields is a signed
counter, which (in the case of false deletions) may go negative.

Since all $n$ ID's in $U$ can be represented with $O(\log n)$ bits,
their sum can also be represented with $O(\log n)$ bits. Thus, the
space needed for $B$ and $C$ is $O(m\log n)=O(d\log n\log (1/\epsilon))$ bits.

\subsection{Updating an Invertible Bloom Filter}
We process updates for the invertible Bloom filter as follows.

\begin{itemize}
\item[]
{\Insert} $x$:
\begin{algorithmic}[100]
\STATE increment {\Count}
\FOR {$i=1,\ldots,k$}
\STATE increment $B[h_i(x)].\Count$
\STATE add $x$ to $B[h_i(x)].\Idsum$
\STATE add $g(x)$ to $B[h_i(x)].\Hashsum$
\ENDFOR
\FOR {$i=1,2$}
\STATE increment $C[f_i(x)].\Count$
\STATE add $x$ to $C[f_i(x)].\Idsum$
\STATE add $g(x)$ to $C[f_i(x)].\Hashsum$
\ENDFOR
\end{algorithmic}
\item[]
{\Delete} $x$:
\begin{algorithmic}[100]
\STATE decrement {\Count}
\FOR {$i=1,\ldots,k$}
\STATE decrement $B[h_i(x)].\Count$
\STATE subtract $x$ from $B[h_i(x)].\Idsum$
\STATE subtract $g(x)$ from $B[h_i(x)].\Hashsum$
\ENDFOR
\FOR {$i=1,2$}
\STATE decrement $C[f_i(x)].\Count$
\STATE subtract $x$ from $C[f_i(x)].\Idsum$
\STATE subtract $g(x)$ from $C[f_i(x)].\Hashsum$
\ENDFOR
\end{algorithmic}
\end{itemize}

That is, to insert $x$, we go to each cell that $x$ maps to and
increment its {\Count} field, add $x$ to its {\Idsum} field, and add
$g(x)$ to its {\Hashsum} field.
Thus, the methods for element insertion is fairly
straightforward. 
Deletion is similarly easy, in that 
we simply decrement counts and subtract out the appropriate
summands to reverse the insertion operation.
These operations are illustrated in Figure~\ref{fig:indel}.

\subsection{Listing the Contents of an Invertible Bloom Filter}
\begin{figure}[t]
\centering\includegraphics[scale=0.9]{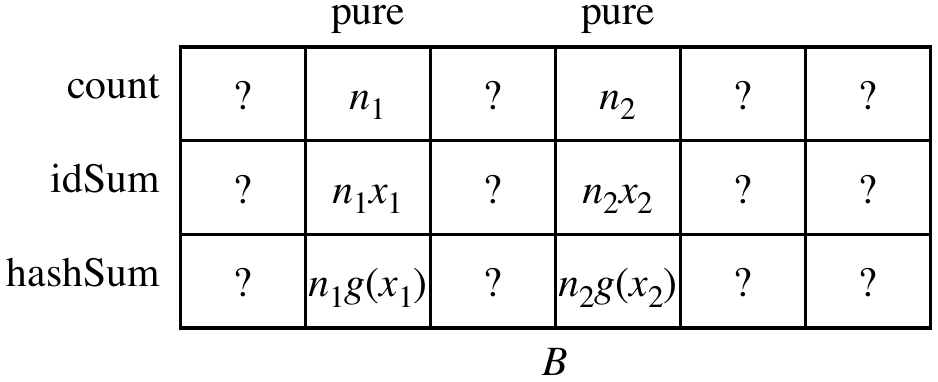}
\caption{Pure cells of $B$ allow us to recover the identity of their items and (using the $\Hashsum$ field) to verify their purity with high probability.}
\label{fig:purity}
\end{figure}

Our method for performing the {\LS} operation is a bit more involved than the insert and delete operations. The basic idea is that some cells of $B$ are likely to be
\emph{pure}, that is, to have values that have been affected by only a single
item (Figure~\ref{fig:purity}). If we can find a pure cell, we can recover the identity of its item by
dividing its $\Idsum$ by its $\Count$. Once a single item and its count are
known, we can remove that item from the database 
and continue until all items have been found. 

The difficulty with this approach is in finding the pure cells. Because
of the possibility of multiple insertions and false deletions, we
cannot simply test whether $\Count$ is one: some pure cells may have
larger counts (i.e., have multiple copies of the same value), 
and some impure cells may have a count equal to one (e.g., because of 
two insertions of a value $x$ followed by a false deletion of a value
$y$ that collides with $x$ at this cell).
Instead, to test whether a cell is pure, we use its $\Hashsum$: in a
pure cell, the $\Hashsum$ should equal the $\Count$ times the hash of
the item's identifier, while in a cell that is not pure it is very
unlikely that the $\Hashsum$, $\Idsum$, and $\Count$ fields will match
up in this way.  

The following pseudo-code expresses the decoding
algorithm outlined above.

\begin{itemize}
\item[] {\LS}:
\begin{algorithmic}
\WHILE {$\exists i$, 
       s.~t.~$g(B[i].\Idsum/B[i].\Count)=B[i].\Hashsum/B[i].\Count$}
\IF[this is a good element] {$B[i].\Count > 0$}
\STATE Push $x=B[i].\Idsum/B[i].\Count$ onto an output stack $O$.
\STATE Delete all $B[i].\Count$ copies of $x$ from $B$ and $C$ 
(using a method similar to {\Delete} $x$ above)
\ELSE[this is a false delete]
\STATE Back out all $-B[i].\Count$ falsely-removed 
copies of $x$ from $B$ and $C$ 
(using a method similar to {\Insert} $x$ above)
\ENDIF
\ENDWHILE
\IF {$\Count = 0$}
\STATE Output the elements in the output stack and insert each
element back into $B$ and $C$.
\ELSE[we have mutually-conflicting elements in $B$]
\STATE Repeat the above while loop, but do the tests using $C$ instead of $B$.
\STATE Output the elements in the output stack, $O$, and insert each
element back into $B$ and $C$.
\ENDIF
\end{algorithmic}
\end{itemize}
There is a slight chance that this algorithm fails.
For example, 
we could have two or more items colliding in a cell of $B$, but we
could nevertheless have the condition,
$g(B[i].\Idsum/B[i].\Count)=B[i].\Hashsum/B[i].\Count$,
satisfied (and similarly for $C$ in the second while loop).
Fortunately, since $g$ is a random function from $[0,n]$ to
$[0,n^2]$, such an event occurs with probability at most $1/n^2$; hence,
over the entire algorithm we can assume, with high probability, that
it never occurs (since $d<<n$).
More troubling is the possibility that,
even after using the fallback array, $C$, 
to find and enumerate elements in the invertible Bloom filter
(in the second while loop),
we might still have some mutually-conflicting elements in $C$.
\ifFull
That is, we would have ${\Count}>0$, even after the second while loop.
Let us therefore analyze this probability of failure for the {\LS} algorithm,
beginning with the first while loop.
\fi

\begin{lemma}
\label{lemma:1}
If the number of elements in $S$, which were inserted but not
deleted, plus the number of false elements negatively indicated in
$S$, corresponding to items deleted but not inserted, is at most $d$,
then the first while loop will remove all but $\epsilon d$ such elements 
from $S$ with probability $1-\epsilon/2$, for $\epsilon<1/4$.
\end{lemma}
\begin{IEEEproof}
\ifFull
It is sufficient for us to show that, with probability $1-\epsilon/2$,
for all but $\epsilon d$
elements $x$ in $S$, there is a cell in $B$ such that that $x$
is the only element in $S$ mapping to that cell.
Let us define the constants so
that each of the $d$ elements in $B$ map to most $k=\log (1/\epsilon)$ 
distinct cells, and the size of $B$ is $4dk$,
which implies that
the probability of a collision at any cell is at most $1/4$.
Thus, the probability that any element $x$ collides with other
elements in each of the cells it gets mapped to is at most $1/4^k$.
That is, 
we can bound 
the number of elements to remain after the first while loop 
using a sum of independent $0-1$ random variables
that has expectation at most $\epsilon^2 d$.
Using this fact, we can use
a Chernoff bound (e.g., see~\cite{mr-ra-95}) to show that
the number of such elements is at most $\epsilon d$ with probability
at least $1-\epsilon/2$.
\else
Omitted due to space limitations.
\fi
\end{IEEEproof}

\begin{figure}[t]
\centering\includegraphics[scale=0.9]{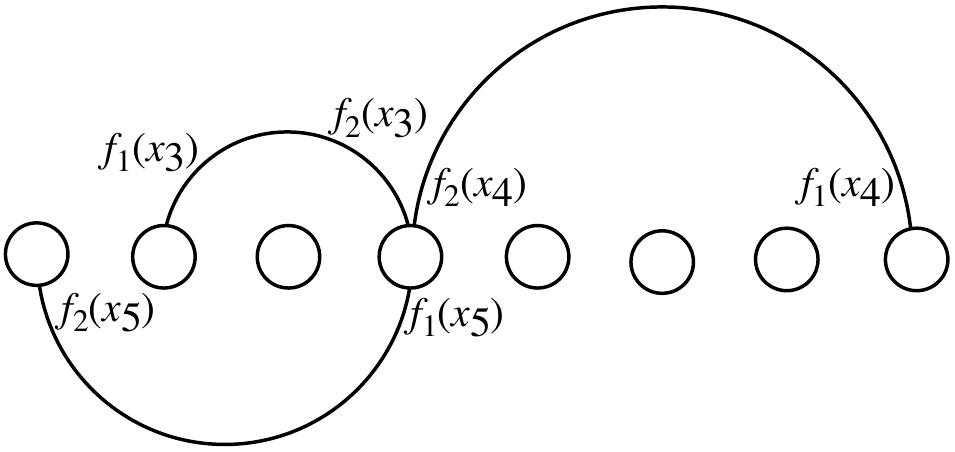}
\caption{A highly sparse random graph in which the vertices represent cells in $C$ and the edges connect cells $f_1(x_i)$ and $f_2(x_i)$ for each remaining element $x_i$. Degree-one vertices of this graph form pure cells in $C$, so if the graph has no cycles it may be uniquely decoded.}
\label{fig:cgraph}
\end{figure}

Let us assume, therefore, that at most $\epsilon d$ elements (true
and/or false) remain in $S$ after the first while loop.
Let us suppose further that each is mapped to two distinct cells in
$C$ (the probability there is any such self-collision among the
remaining elements in $C$ is at most $\epsilon d/4dk\le \epsilon/4$).
We can envision each cell in $C$ as forming a
vertex in a graph, and each selected pair of cells as forming an edge
in the graph (Figure~\ref{fig:cgraph}); thus our data can be modeled as a random multigraph
with $x\le \epsilon d$ edges and 
$y=4dk\ge 8d$ vertices. 
Thus, it is a very sparse graph.
Let $c=y/x\ge 8/\epsilon$.

Two types of bad event could prevent us from decoding
the data remaining in $C$ after the first loop.
First, two items could map to the same pair of
cells, so that our multigraph is not a simple graph. There are
$x(x-1)/2$ pairs of items, and each two items collide with
probability $2/(y(y-1))$, so the expected number of collisions of this
type is $x(x-1)/(y(y-1))$, roughly $1/c^2$.
Second, the graph may be simple but may contain a cycle. 
As shown by Pittel \cite[Exercise 8, p. 122]{b-rg-85}, the
expected number of vertices in cyclic components of a random graph of
this size is bounded by $\sum_{k=3}^{\infty} kc^-k=O(1/c^3)$.
Therefore, the expected number of events of either type,
and the probability that there exists an event of either type, is $O(1/c^2)$. Choosing $c=O(\sqrt{1/\epsilon})$ is 
sufficient to show that we will fail in the second while loop with
probability at most $\epsilon/4$.

\begin{theorem}
If the number of elements in $S$, which were inserted but not
deleted, plus the number of false elements negatively indicated in
$S$, which correspond to items deleted but not inserted, is at most $d$,
then the above algorithm correctly answers a {\LS} query with
probability at least $1-\epsilon$, where $\epsilon<1/4$.
\end{theorem}

\begin{figure}[hbt]
\centering\includegraphics[width=3in]{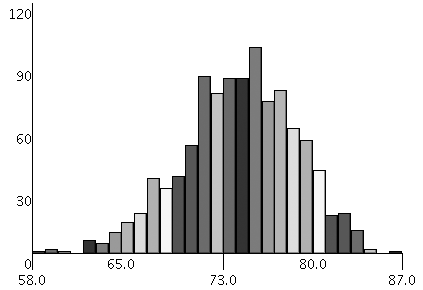}
\caption{Frequencies of saturation points for $B$ used alone. The
mean is 74.8 and the standard deviation is 4.4.}
\label{fig:histogram1}
\end{figure}

\begin{figure}[hbt]
\centering\includegraphics[width=3in]{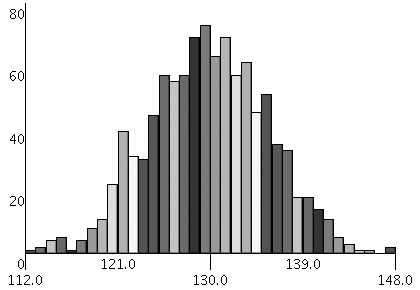}
\caption{Frequencies of saturation points for $B$ and $C$ used together.
The mean is 130.3 and the standard deviation is 5.7.}
\label{fig:histogram2}
\end{figure}

To get a handle on the real-world performance of the invertible Bloom
filter, we implemented an instance of the table $B$, with four
random hash functions and capacity of 101 cells.
The four hash functions and the functions $f_1$ and $f_2$ were
implemented using the SHA-1 cryptographic hash function, modulo 101,
and the hash function $g$ was implemented using the SHA-1
function, modulo 10211.
We then inserted as many elements as possible such that we could
still perform the {\LS} operation (without resorting to the backup table $C$).
We implemented the {\Count} and {\Idsum} fields using 16-bit
integers, and we implemented the {\Hashsum} field using a 32-bit
integer.
We did one set of experiments with 
the table $B$ used alone and another set of experiments with the
table used in conjunction with the table $C$.
In both cases, we searched for clean elements as described above, but
also added a ``sanity'' check that tests that each clean element being
listed in a {\LS} operation actually maps to the location that
revealed this clean element.
We performed 1000 random trials of each set of experiments, 
and we show a histogram of the maximum sizes of feasible inversions,
for both sets,
with the results for $B$ used alone shown
in Figure~\ref{fig:histogram1}.
and those for $B$ and $C$ used together in Figure~\ref{fig:histogram2}.
Clearly, the use of the backup table, $C$, significantly extends the
ability of the invertible Bloom filter to recover a set.

\section{Conclusion and Future Directions}
In this paper, we study the straggler identification problem for data
streams, showing that small sublinear-space indexing schemes exist for 
performing straggler detection.
Another way of viewing this problem is that we desire a database indexing scheme
that can represent a dynamic set using a compact structure, $D$.
As the database $D$ fills to be of size as large as $n$, the cells of $D$
can ``overflow'' and we lose the ability to list the contents of $D$.
But as items are removed from $D$, we eventually get to a point where we can
enumerate the contents of $D$ again.

Our deterministic solution uses $O(d\log n)$ bits to represent $D$, where $d$
is a parameter indicating an upper bound on the number of stragglers we
expect to exist at the time when we wish to enumerate the contents of $D$.
We observe that this deterministic solution cannot tolerate redundant
insertions or false deletions, but this requirement is justified by our
negativity result for any deterministic solution to the straggler
identification problem.
Our randomize solution, on the hand, which introduces the invertible Bloom
filter, can tolerate both redundant insertions and false deletions, provided
there are not too many of them.

In all our solutions,
we assume we have an upper bound, $d$, on the size of $D$ at the time we wish
to perform enumerations of its contents.
One direction of future study, then, is to reduce this requirement of
knowledge of an upper bound $d$, for example, for insertion-deletion
sequences that belong to certain probabilistic distributions.

\subsection*{Acknowledgments}
We would like to thank Dan Hirschberg for several helpful discussions. 
We are also grateful to an anonymous reviewer for suggesting the multicast 
application. 
a preliminary and abridged version of this paper was 
presented~\cite{eg-sesir-07}
at the 
10th Workshop on Algorithms and Data Structures, 
Halifax, Nova Scotia, 2007. 
The authors' research was supported in part by NSF grant
0830403 and by the Office of Naval Research under grant
N00014-08-1-1015.

\bibliography{geom,extra,cgt,goodrich}
\ifIEEE
\else
\bibliographystyle{abbrv}
\fi

\end{document}